\documentclass[a4paper,10pt,twoside]{cpc-hepnp}
\usepackage{CJK,upgreek,fancyhdr}
\usepackage{multicol}
\usepackage{graphicx}
\usepackage{booktabs}
\usepackage{amssymb,bm,mathrsfs,bbm,amscd}
\usepackage[tbtags]{amsmath}
\usepackage{lastpage}

\usepackage{color}
\usepackage[normalem]{ulem}

\usepackage{slashed}
\usepackage{tabularx}

\usepackage{amssymb}
\usepackage{amsmath}
\usepackage{amsfonts}
\usepackage{epsfig}
\usepackage{graphicx}
\usepackage{tabularx}
\usepackage{latexsym}
\usepackage{subfigure}
\usepackage{verbatim}
\usepackage{color}
\usepackage{braket}
\usepackage{multirow}
\usepackage{dcolumn}
\usepackage{placeins}
\usepackage{epstopdf}
\usepackage{wrapfig}
\usepackage{hyperref}
\usepackage{booktabs}
\usepackage{simplewick}
\usepackage[normalem]{ulem}
\usepackage[toc,page]{appendix}
\def\gev{\mathrm{GeV}}

\renewcommand\sout{\bgroup \color{red} \ULdepth=-.5ex \ULset}

\def\eq{\begin{eqnarray}}
\def\en{\end{eqnarray}}

\begin{document}
\begin{CJK*}{GB}{gbsn}

\fancyhead[c]{\small Chinese Physics C~~~Vol. xx, No. x (201x) xxxxxx}
\fancyfoot[C]{\small xxxxxx-\thepage}

\footnotetext[0]

\title{$\rho$ meson impact parameter distributions\thanks{Supported by the National Natural Science Foundation of China under Grant No. 11475192, by the fund provided to the Sino-German CRC 110 "Symmetries and the Emergence of Structure in QCD" project by the NSFC under Grant No. 11621131001,  and the Key Research Program of Frontier Sciences,  CAS, Grant No. Y7292610K1.}}

\author{%
      Bao-Dong Sun$^{1,2;1)}$\email{sunbd@ihep.ac.cn}
\quad Yu-Bing Dong$^{1,2,3}$
}
\maketitle

\address{%
$^1$Institute of High Energy Physics, Chinese Academy of Sciences, Beijing 100049, P. R. China\\
$^2$School of Physics, University of Chinese Academy of Sciences, Beijing 100049, P. R. China\\
$^3$Theoretical Physics Center for Science Facilities (TPCSF), CAS, Beijing 100049, P. R. China\\
}

\begin{abstract}
In this paper, the $\rho$ meson impact parameter dependent parton distributions and the impact parameter dependent
form factors are introduced and discussed. By employing a Gaussian form wave packet, we calculate the impact parameter distributions of the $\rho$ meson based on a light-cone constituent quark model.
\end{abstract}



\footnotetext[0]{\hspace*{-3mm}\raisebox{0.3ex}{$\scriptstyle\copyright$}2018
Chinese Physical Society and the Institute of High Energy Physics
of the Chinese Academy of Sciences and the Institute
of Modern Physics of the Chinese Academy of Sciences and IOP Publishing Ltd}%

\begin{multicols}{2}

\section{Introduction}\label{sec:introduction}

It is common believed that the usual parton distributions (PDFs) can only give the longitudinal information of a hadron target in the deep inelastic scattering (DIS) processes, while the generalized parton distributions (GPDs) have the promising ability to shade light on the transverse information, which gives rise to the idea of ``quark/gluon imaging" of hadrons~\cite{Marukyan2015}. Moreover, the impact parameter distributions (IPDs),
obtained by the Fourier transform of GPDs with respect to the transverse momentum transfer, may show some
information about the transverse impact space position of partons~\cite{Burkardt:2002hr}. This impact parameter representation is useful in processes such as high-energy scattering and hard processes~\cite{Diehl2003}. It is also argued that, in position space, IPDs play a similar role to the charge distributions, and are, thus, 
very promising for understanding the hadron internal structures. \\

As we know, $G_C(Q^2)$ is the form factor of the conserved local current, and is thus independent of the renormalization scale $\mu$. It can be obtained through the sum rules from GPDs, which by definition are probed in hard processes~\cite{Diehl2003}. In the case of Fourier transforms of GPDs, Burkardt pointed out that, when $\xi=0$, the Fourier transforms of GPDs have the interpretation of a density of partons with longitudinal momentum fraction $x$, localized at ${\bf b}_\perp$ relative to the transverse center in the impact parameter space, which is allowed by the Heisenberg uncertainty principle~\cite{Burkardt:2000za,Diehl:2002he}. Due to the significance of the form factors in the impact parameter space, many theoretical works have been devoted to study the IPDs of pions, kaons and nucleons~\cite{Miller2009,Miller:2007uy,Miller:2010nz,Miller:2009qu,Miller:2010tz,
Carmignotto:2014rqa,Nam:2011yw,Kumar:2015yta,Diehl:2002he,Dalley2003Jul21,Broniowski2003}.  \\

It should be mentioned that our recent work~\cite{Sun:2017gtz} gave a discussion of the $\rho$ meson unpolarized GPDs in  momentum space with a Light-Cone Constituent Quark Model (LCCQM).
The form factors and some other low-energy observables of the $\rho$ meson were calculated and our numerical results agreed with the previous publications and some experimental data~\cite{Krutov:2018mbu}.
In the literature, the constituent quark model is also used to describe the form factors of pions, nucleons, deuterons, etc.~\cite{Frederico:2009fk,Sun:2016ncc,Dahiya:2017fmp}.
Moreover, the contributions from the valence and non-valence regimes to the form factors and generalized parton distributions were discussed and analyzed in detail.
In addition, the reduced matrix elements, which are the moments of the DIS structure functions, were also estimated and the obtained values were compatible with the available lattice calculation at the same scale ratio~\cite{Best1997}.
In general, our numerical results for the unpolarized GPDs~\cite{Sun:2017gtz} were reasonable and satisfying. Therefore, in this work, we extend the phenomenological model to study the IPDs of the $\rho$ meson and to calculate the impact parameter dependent PDFs of $q(x,{\bf b}_\perp)$ and
$q({\bf b}_\perp)$ and the form factors of  $q^{C,M,Q}(x,{\bf b}_\perp)$ and $q^{C,M,Q}({\bf b}_\perp)$. \\

The paper is organized as follows. In Section~\ref{Impact_parameter_dependent_PDF}, the framework of the impact
parameter dependent PDFs is presented. In Section~\ref{sec:Wave_packets}, we discuss the wave packets and the
cutoff for the numerical calculation. The definitions of the impact parameter dependent FFs are
given in Section~\ref{sec:impact_parameter_space_FFs}. Our numerical results for the PDFs and FFs in the impact
parameter space are shown in Section~\ref{sec:Results}, and Section~\ref{sec:summary} gives a short
summary and conclusion.\\

\section{Impact parameter dependent PDFs}
\label{Impact_parameter_dependent_PDF}

When considering the nucleon GPDs without helicity flip, Burkardt~\cite{Burkardt2003} identifies the Fourier transform of its GPD $H_q(x,\xi=0,-{\bf \Delta}_\perp^2)$ w.r.t. $-{\bf \Delta}_\perp^2$ as a distribution of partons in the transverse plane, i.e., the probability of finding a quark with longitudinal momentum fraction $x$ and at transverse impact space position ${\bf b}_\perp$.
The impact parameter dependent PDF for a nucleon (a spin-1/2 target), given by Ref.~\cite{Burkardt2003}, reads
\end{multicols}
\eq
q_N(x,{\bf b}_\perp)
&=& \left|{\cal N}\right|^2
\int \frac{d^2{\bf p}_\perp}{(2\pi)^2}
\int \frac{d^2{\bf p}_\perp^\prime}{(2\pi)^2}\times
\langle p^+, {\bf p}^\prime_\perp, \lambda | \left[
\int\frac{dz^-}{4\pi} \bar{q}(-\frac{z^-}{2}, {\bf b}_\perp) \gamma^+ q(\frac{z^-}{2}, {\bf b}_\perp)
e^{-\imath x p^+ z^-}
\right] | p^+, {\bf p}_\perp, \lambda \rangle \nonumber\\
&=& \left|{\cal N}\right|^2
\int \frac{d^2{\bf p}_\perp}{(2\pi)^2}
\int \frac{d^2{\bf p}_\perp^\prime}{(2\pi)^2}
H_q(x,\xi=0,-\left({\bf p}_\perp-{\bf p}_\perp^\prime \right)^2)
e^{i{{\bf b}_\perp} \cdot ({\bf p}_\perp-{\bf p}_\perp^\prime)}
\nonumber\\
&=&
\int \frac{d^2{\bf \Delta}_\perp}{(2\pi)^2}
H_q(x,0,-{\bf \Delta}_\perp^2)
e^{-i{\bf b_\perp} \cdot {\bf \Delta}_\perp}
\nonumber \\
&=&
\int_0^{\infty} \frac{\Delta_\perp d
\Delta_\perp}{2\pi} J_0 (b \Delta_\perp) H_q(x,0,-\Delta_\perp^2) \nonumber \\
&=& q_N(x,{b}),
\label{eq:result1}
\en
\begin{multicols}{2}
where the normalization factor ${\cal N}$ satisfies $\left|{\cal N}\right|^2\int\frac{d{\bf p_\perp}}{(2\pi)^2}=1$,
and $\Delta_\perp=|{\bf \Delta}_\perp|=\sqrt{\Delta_x+\Delta_y}$ and $b=|{\bf b}_\perp|=\sqrt{b_x+b_y}$. Cylindrical symmetry is applied in the last but one step and $J_0$ is the Bessel function of the first kind $J_\nu(z)$ with $\nu=0$. The parton distribution depends on transverse impact space position ${\bf b}_\perp$ only through its norm $b$ being the consequence of the longitudinal polarization.
In the third step the integral turns to the total and transverse momentum transfer, i.e., ${d^2{\bf p}_\perp}{d^2{\bf p}_\perp^\prime}={d^2{\bf \Delta}_\perp}{d^2{\bf P}_\perp}$, with ${\bf \Delta}_\perp={\bf p}_\perp^\prime-{\bf p}_\perp$ and $\bf P_\perp= (\bf p^\prime_\perp + \bf p_\perp)/2$, and using the fact that GPD $H$ is independent of total transverse momentum $\bf P_\perp$.
Ignoring the helicity flip, the spin projection $\lambda$ can be dropped. In the forward limit, namely $\xi=0$, we have $t=(p'-p)^2=-{\bf \Delta}_\perp^2$. \\

Note that Hoodbhoy~\cite{Hoodbhoy:1988am} has already pointed out the DIS structure function
$F_1$, $F_2$, $g_1$, and $g_2$ of spin-1 targets can be precisely measured in the same way as that
of spin-1/2 targets. Analogous to the fact that the structure function $F_1$ connects to GPD $H_q$ for spin-1/2 targets, we simply assume $F_1$ connects to the GPD $H_1^q$ for spin-1 targets as well.
As shown by Eqs.~(37$\sim$39) in Ref.~\cite{Sun:2017gtz}, the isospin combination implies that
\eq
\int_{-1}^1 dx \; H_i^{u} (x,\xi,t) = \int_{-1}^1 dx \; H_i^{I=1} (x,\xi,t) \ .
\en
Hereafter we omit the label of quark flavor $u$ and isospin $I=1$ for simplicity.
Due to the similar roles of $H_q$ and $H_1$, we introduce the impact parameter dependent PDF for spin-1 targets (for the $u$ quark),
\eq
q(x,b)
&=&
\int_0^{\infty} \frac{\Delta_\perp d
\Delta_\perp}{2\pi} J_0 (b \Delta_\perp) H_1(x,0,-\Delta_\perp^2) \ ,
\label{eq:result2}
\en
One can further define the total parton distribution in the impact parameter space as 
\eq
\hspace{-10mm} q(b)
&=&\int_0^1 dx \; q(x,b) \ . 
\label{eq:result3}
\en 
\par\noindent
Notice that $\int d^2{\bf b}_\perp \; q(x,{ b})=H_1(x,0,0)$, which is equal to the usual PDF $q(x)$ in the forward limit $t=\Delta^2\rightarrow0$. Therefore, $q(x,b)$, the Fourier transform of the GPD $H_1(x,\xi=0,-\Delta_\perp^2)$ w.r.t. $- \Delta_\perp^2$, can be identified, in analogy to the nucleon case, with the probability of finding a quark
with longitudinal momentum fraction $x$ and  transverse impact space position ${\bf b}_\perp$ in the $\rho$ meson. \\

It should be emphasized that in Ref.~\cite{Burkardt:2002hr}, the nucleon impact parameter dependent PDF $q_N$ was proved to  satisfy the positive constraints for the so-called ``good" quark field. In our model calculation, the phenomenological vertexes (see Eq.~(24) in Ref.~\cite{Sun:2017gtz}) involve the loop momentum ($k$), and the form of the vertexes is fixed according to the constraints from isospin symmetry. Our sophisticated model cannot simply reproduce the procedure of Ref.~\cite{Burkardt:2002hr} to fold the correlation function into a norm of a quantity (see Eq.~(23) of Ref.~\cite{Burkardt:2002hr}). Therefore, the positive constraint for $q(x,b)$ with a realistic model calculation needs to be proven further.\\

\section{Wave packets}
\label{sec:Wave_packets}

The Fourier transform of a plane wave is not well defined, thus, one can start with the wave packets instead of the plane wave. In the non-relativistic limit, the Fourier transform of the charge form factor $G_C(Q^2)$ can
be interpreted as the charge distribution in the transverse direction. In other words, as long as the wave packets peak sharply at some point in position space, by taking the non-relativistic limit, the Fourier transform of the charge distribution equals the form factor. By the way, a Gaussian weighting factor was also adopted in a recent lattice QCD calculation~\cite{Chen:2017lnm}, in order to suppress the unphysical oscillatory behaviour. The oscillation
is due to the finite lattice size and nucleon momentum. The result in the small Bjorken $x$($<0.3$) region is changed
by weighting. In Ref.~\cite{Pire:2002ut}, the Gaussian ansatz is also applied to shape the hadron when calculating generalized distribution amplitudes of the pion pair production process.
\\

Moreover, as pointed out by Burkardt~\cite{Burkardt:2000za,Burkardt:2002hr}, the interpretation of the Fourier transform of the form factor as the charge distribution may receive relativistic corrections in the rest frame. However, such a problem may disappear in either Breit frame or infinite momentum frame (IMF). In the relativistic case, the transform receives relativistic corrections when the wave packet is localized with a size smaller than the Compton wavelength
of the system. In the IMF, the relativistic correction can be managed to be very small, and therefore, the wave packet
does not change the interpretation, as long as the wave packets are set slowly varying w.r.t. $\bf \Delta_\perp$. To be specific, the width of the wave packets must be much larger than a typical QCD scale $\Lambda_{QCD}$ ($\sim0.23~\gev$). For a Gaussian form wave packet, one gets $\sigma\ll 1/\Lambda_{QCD} \sim 3/M$,  with $M$ being the $\rho$ meson mass. \\

On the other hand, as Diehl~\cite{Diehl:2002he} has discussed, a real hadron is an extended object and is smeared out by an amount $\sigma$. From the experimental viewpoint, there is a largest measured value $|t|_{\text{max}}$ and thus there is the accuracy of the measurement $\sigma\sim(|t|_{\text{max}})^{-1/2}$. According to the observations and to the limit of the effect from unmeasured values of $t$, a Gaussian form wave packet can also be reasonably introduced. Thus we have
\eq
&&\int \frac{d^2{\bf p}_\perp dp^+}{(2\pi)^2 p^+} p^+\delta(p^+-p^+_0) G({\bf p}_\perp, \frac{1}{\sigma^2}) | p,\lambda \rangle
\nonumber \\
&&\sim \;
\int \frac{d^2{\bf p}_\perp}{(2\pi)^2} \text{exp}\left( -\frac{{\bf p}^2_\perp \sigma^2}{2}  \right)| p^+, {\bf p}_\perp, \lambda \rangle \ ,
\label{eq:wp}
\en
where $G({\bf p}_\perp, 1/\sigma^2)=\text{exp}( -{\bf p}^2_\perp \sigma^2/2 )$ and
the mixed state is modified to be
\eq \label{eq:state}
| p^+, {\bf b}_\perp, \lambda \rangle_\sigma
&=&
{\cal N}_{\sigma} \int \frac{d^2{\bf p}_\perp}{(2\pi)^2}
e^{-\imath {\bf b}_\perp \cdot {\bf p}} G({\bf p}_\perp, \frac{1}{\sigma^2})
| p^+, {\bf p}_\perp, \lambda \rangle
\nonumber \\
&\stackrel{\sigma\rightarrow0}{=} &| p^+, {\bf b}_\perp, \lambda \rangle
\ ,
\en
where the normalization factor ${\cal N}_{\sigma}$ satisfies $\left|{\cal N}_{\sigma}\right|^2\int\frac{d{\bf p_\perp}}{(2\pi)^2}=1$ and $\lim_{\sigma\rightarrow0} {\cal N}_{\sigma}={\cal N} $. Note that our normalization of states is different from that in Ref.~\cite{Diehl:2002he}.
This action will add two Gaussian functions in the expression, $G({\bf p}_\perp, \frac{1}{\sigma^2})$ and $G({\bf p}^\prime_\perp, \frac{1}{\sigma^2})$, into the definition of $q(x,{b})$
(see eq. (2)). We can still change variables to remove the dependence of ${\bf P}_\perp$, which leaves only one $G({\bf \Delta}_\perp, \frac{1}{\sigma^2})$.
Consequently, the definition of the impact parameter dependent PDF is modified to be 
\begin{eqnarray}
q_{\sigma}( x,{ b} )
&=&
\int_0^{\infty} \frac{\Delta_\perp d
\Delta_\perp}{2\pi}
J_0 (b \Delta_\perp) G({\bf \Delta}_\perp,\frac{2}{\sigma^2})
H_1(x,0,-\Delta_\perp^2) \nonumber \\
&=&
\int_0^{\infty} \frac{\Delta_\perp d
\Delta_\perp}{2\pi}
J_0 (b \Delta_\perp) e^{-{{\Delta}_\perp^2}{\sigma^2}/4}
H_1(x,0,-\Delta_\perp^2) \ , \nonumber \\
\label{fb}
\end{eqnarray} 
and 
\begin{eqnarray}
q_{\sigma}({b} )
&=&
\int_0^1 dx \; q_{\sigma}( x,{ b} ) \ .
\label{fb2}
\end{eqnarray} 

Reference~\cite{Diehl:2002he} also argued that in order to give a well-defined (positive, or without sign flip) longitudinal momentum $p^3$, $|{\bf p}_\perp|\ll p^+$ is required. However, as one can see in Eq.~(\ref{eq:wp}), ${\bf p}_\perp$ and $p^+$ are separated in the wave packet and thus this requirement actually does not affect the result of the integrals. This can also be seen from the property of GPDs. In the forward limit, $H(x,0,-{\Delta}_\perp^2)$ is not affected by this requirement either. Moreover, Ref.~\cite{Brodsky:1997de} emphasized that since the longitudinal momentum is $p^+$ in the front form, one needs not to go to infinite momentum along the moving direction, and not to impose the constraint on the $p^3$ component either. \\

According to the above discussions, the relation $\sigma\sim(|t|_{\text{max}})^{-1/2}$ inspires us to introduce a
cutoff ($\Delta_0$) of the momentum transfer in the integral as well
\begin{eqnarray}
\hspace{-5mm} q( x,{b},\Delta_0 )
&=&
\int_0^{\Delta_0} \frac{\Delta_\perp d
\Delta_\perp}{2\pi}
J_0 (b \Delta_\perp) H_1(x,0,-{\Delta}_\perp^2) \ ,
\label{fxbcut}
\end{eqnarray}
and 
\begin{eqnarray}
\hspace{-10mm} q({b},\Delta_0 )
&=&
\int_0^1 dx \; q( x,{b},\Delta_0 ) \ . 
\label{fbcut}
\end{eqnarray} 
This assumption is supported by a comparison between the results of the integrals with a wave packet, $q_\sigma(b)$ (width $\sigma\sim1/\Delta_0$) and the one with a cutoff $q(b,\Delta_0)$. This will be shown in Section~\ref{sec:Results}.  \\

\section{Impact parameter dependent FFs}
\label{sec:impact_parameter_space_FFs}

We emphasize that the unpolarized impact parameter dependent PDFs are proposed to describe the transverse distribution of unpolarized partons in an unpolarized target. As shown in previous sections, the IPDs can be  obtained through Fourier transform of the unpolarized GPD $H_1$. We notice that the conventional charge,
magnetic dipole and quadrupole FFs are the integrals of the linear combination of $H_i$. This motivates us to explore the possibility of obtain the IPDs with respect to the three FFs. The sum rules relating to the GPDs and the
FFs $G_i$ are~\cite{Berger2001} 
\eq\label{eq:sumrule}
\int_{-1}^{1} dx H_i (x,\xi,t) &=& G_i(t) \quad (i=1,2,3) \ , \nonumber \\
\int_{-1}^{1} dx H_i (x,\xi,t) &=& 0  \quad (i=4,5) \ ,
\en 
where $G_i^q$ are the FFs in the decomposition of the local current.
The FFs $G_{C,M,Q}$ can be expressed in terms of $G_{1,2,3}$ as~\cite{Choi2004}
\begin{eqnarray}
G_C(t)&=&G_1(t) + \frac{2}{3}{\eta} G_Q(t) \ , \ \nonumber \\
G_M(t)&=&G_2(t) \ , \ \nonumber \\
G_Q(t)&=&G_1(t) - G_2(t) + (1+\eta)G_3(t)\ , \
\label{eq:Gcmq}
\end{eqnarray}
where $\eta=-t/4M^2$. Together with Eq.~(\ref{eq:sumrule}), one can obtain $G_{C,M,Q}$ directly from GPDs $H_{1,2,3}$. This allows us to bypass the well-known ambiguity of the angular condition~\cite{Melo1997}.
With the above two equations, one can get the relations
\end{multicols}
\eq
G_C(t)&=&\int_{-1}^{1} dx \Big [H_1(x,\xi,t) + \frac{2}{3}{\eta} \left[H_1(x,\xi,t) - H_2(x,\xi,t) + (1+\eta)H_3(x,\xi,t) \right] \Big ]\ , \ \nonumber \\
G_M(t)&=&\int_{-1}^{1} dx H_2(x,\xi,t) \ , \ \nonumber \\
G_Q(t)&=&\int_{-1}^{1} dx \Big [H_1(x,\xi,t) - H_2(x,\xi,t) + (1+\eta)H_3(x,\xi,t)\Big ]\ . \
\en
\begin{multicols}{2}

Notice that by taking $\xi=0$ and $\eta=-t/4M^2={\Delta}_\perp^2/4M^2$,
one can get quantities similar to the integrands in Eq.~(\ref{eq:result1}).
We have the impact parameter dependent FFs
\end{multicols}
\eq
\label{eq:ipFFsGC}
q^C_\sigma(x,{b})
&=&
\int_0^{\infty} \frac{\Delta_\perp d
\Delta_\perp}{2\pi}
J_0 (b \Delta_\perp)
e^{-{{\Delta}_\perp^2}{\sigma^2}/4}
\nonumber \\ &&\times
\Bigg[ H_1(x,0,-{\Delta}_\perp^2) + \frac{2}{3}{\frac{{\Delta}_\perp^2}{4M^2}} \Big[H_1(x,0,-{\Delta}_\perp^2) - H_2(x,0,-{ \Delta}_\perp^2)+ (1+{\frac{{\Delta}_\perp^2}{4M^2}})H_3(x,0,-{\Delta}_\perp^2) \Big] \Bigg] \ , \\\
\label{eq:ipFFsGM}
q^M_\sigma(x,{b})
&=&
\frac{1}{G_M(0)}\int_0^{\infty} \frac{\Delta_\perp d
\Delta_\perp}{2\pi}
J_0 (b \Delta_\perp)
e^{-{{\Delta}_\perp^2}{\sigma^2}/4}
H_2(x,0,-{ \Delta}_\perp^2) \ , \  \\
q^Q_\sigma(x,{b})
&=&
\frac{1}{G_Q(0)}\int_0^{\infty} \frac{\Delta_\perp d
\Delta_\perp}{2\pi}
J_0 (b \Delta_\perp)
e^{-{{\Delta}_\perp^2}{\sigma^2}/4} \nonumber \\ &&\times
\left[H_1(x,0,-{ \Delta}_\perp^2) - H_2(x,0,-{ \Delta}_\perp^2) + (1+{\frac{{ \Delta}_\perp^2}{4M^2}})H_3(x,0,-{ \Delta}_\perp^2) \right] \ , \
\label{eq:ipFFsGQ}
\en
\begin{multicols}{2}
and
\begin{eqnarray}
\hspace{-10mm} q^{C,M,Q}_{\sigma}({ b} )
=
\int_0^1 dx \; q^{C,M,Q}_{\sigma}( x,{ b} ) \ .
\end{eqnarray}

Comparing the impact parameter dependent FFs, Eq.~(\ref{eq:ipFFsGC}), with the impact parameter dependent PDFs, Eq.~(\ref{fb}), we introduce the ``difference" quantities
\end{multicols}
\eq
q^{QC}_\sigma(x,{b})
&=&
\int_0^{\infty} \frac{\Delta_\perp d
\Delta_\perp}{2\pi}
J_0 (b \Delta_\perp)
e^{-{{\Delta}_\perp^2}{\sigma^2}/4}  \nonumber \\ &&\times
\left(\frac{2}{3}{\frac{{\Delta}_\perp^2}{4M^2}}\right)
\Bigg[H_1(x,0,-{\Delta}_\perp^2) - H_2(x,0,-{\Delta}_\perp^2) 
+ (1+{\frac{{\Delta}_\perp^2}{4M^2}})H_3(x,0,-{\Delta}_\perp^2) \Bigg] \ ,  \\
q^{QC}_{\sigma}({ b} )
&=& \int_0^1 dx \; q^{QC}_{\sigma}( x,{b} ) \ ,
\en
\begin{multicols}{2}
which receive the contribution from the quadrupole moment.
The ``difference" quantities satisfy
\eq
\label{eq:sumCQC}
q^{QC}_{\sigma}( x,{ b} ) &=& q^{C}_{\sigma}( x,{b} ) - q_{\sigma}( x,{ b} ), \; \nonumber \\
q^{QC}_{\sigma}( { b} ) &=& q^{C}_{\sigma}( { b} ) - q_{\sigma}( { b} ).
\en 
It is clear that the impact parameter dependent PDFs relate to the impact parameter dependent FFs and
\eq
\label{eq:sumCMQ}
\int_0^1 dx \int_{-\infty}^{\infty} d^2{\bf b} \; q^{C,M,Q}_\sigma(x,{ b})=1 \; \ . 
\en
Thus, it is possible to interpret $q^{C}_\sigma$, $q^{M}_\sigma$ and $q^{Q}_\sigma$ as the percentage of the
contributions to the charge (normalized to 1), magnetic dipole $\mu_{\rho}$ and quadrupole moment $Q_\rho$ respectively,
from the parton with the longitudinal momentum fraction $x$ and transverse impact space position ${\bf b}_\perp$.
\\

\section{Results}\label{sec:Results}

In our previous work~\cite{Sun:2017gtz} with  a light-cone constituent quark model, we took the two model parameters of the constituent mass $m=0.403~\gev$ and regulator mass $m_R=1.61~\gev$, and we calculated the GPDs of the $\rho$ meson.
In our LCCQM, we introduced an effective Lagrangian for the $\rho-q\bar q$ interaction with a phenomenological vertex $\Gamma^u$ and a Bethe-Salpeter amplitude. By integrating the minus component of the quark momentum $k^-$ analytically and rest of the components numerically, we obtained the GPDs and FFs of the $\rho$ meson.

\begin{figure}
\centering
{\hskip -1.5cm}
\subfigure[~$q_\sigma(b)$ with packet width $\sigma=1/2~\gev^{-1}$, $1~\gev^{-1}$, and $2~\gev^{-1}$.]{\label{fig:ipPDF:s}
\begin{minipage}[b]{0.4\textwidth}
\includegraphics[width=1\textwidth]{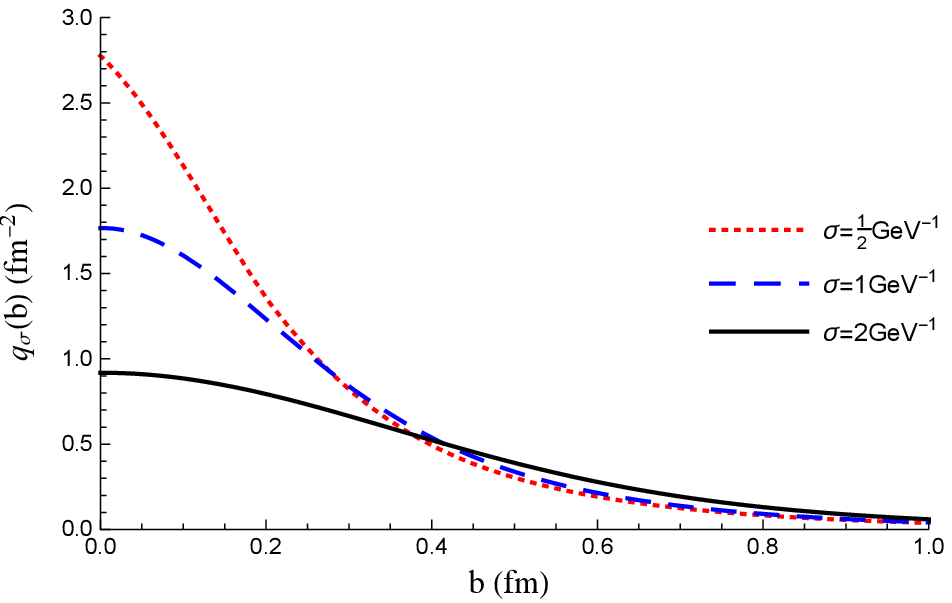}
\end{minipage}
}
{\hskip 1cm}
\subfigure[~$q(b,\Delta_0)$ with cutoff $\Delta_0=1/2~\gev$, $1~\gev$ and $2~\gev$.]{\label{fig:ipPDF:c}
\begin{minipage}[b]{0.4\textwidth}
\includegraphics[width=1\textwidth]{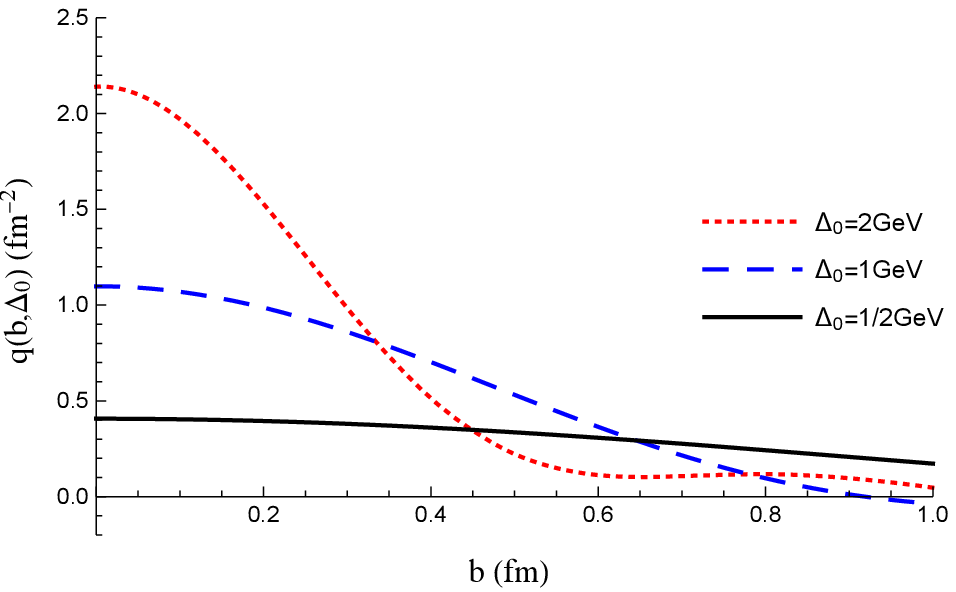}
\end{minipage}
}
\caption{\label{fig:ipPDF}{ \small The impact parameter dependent PDF $q(b)$ with (a) a wave packet and (b) a cutoff on the momentum transfer.
}}
\end{figure}

\begin{figure}
\centering
{\hskip -1.5cm}
\subfigure[~$q_\sigma^C(b)$ with packet width $\sigma=1/2~\gev^{-1}$, $1~\gev^{-1}$, and $2~\gev^{-1}$.]{\label{fig:ipPDFCQC:C}
\begin{minipage}[b]{0.4\textwidth}
\includegraphics[width=1\textwidth]{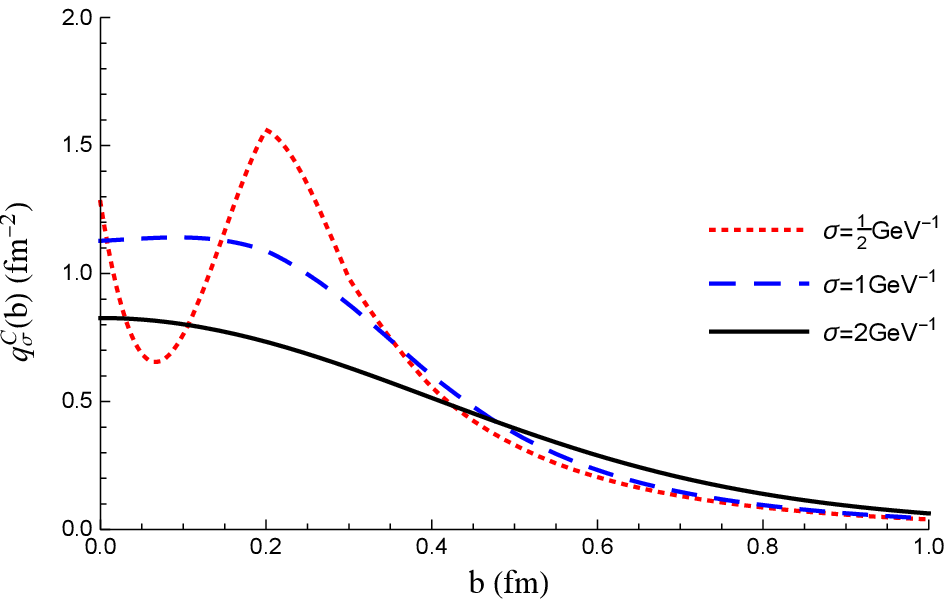}
\end{minipage}
}
{\hskip 1cm}
\subfigure[~$q_\sigma^{QC}(b)$ with packet width $\sigma=1/2~\gev^{-1}$, $1~\gev^{-1}$, and $2~\gev^{-1}$.]{\label{fig:ipPDFCQC:QC}
\begin{minipage}[b]{0.4\textwidth}
\includegraphics[width=1\textwidth]{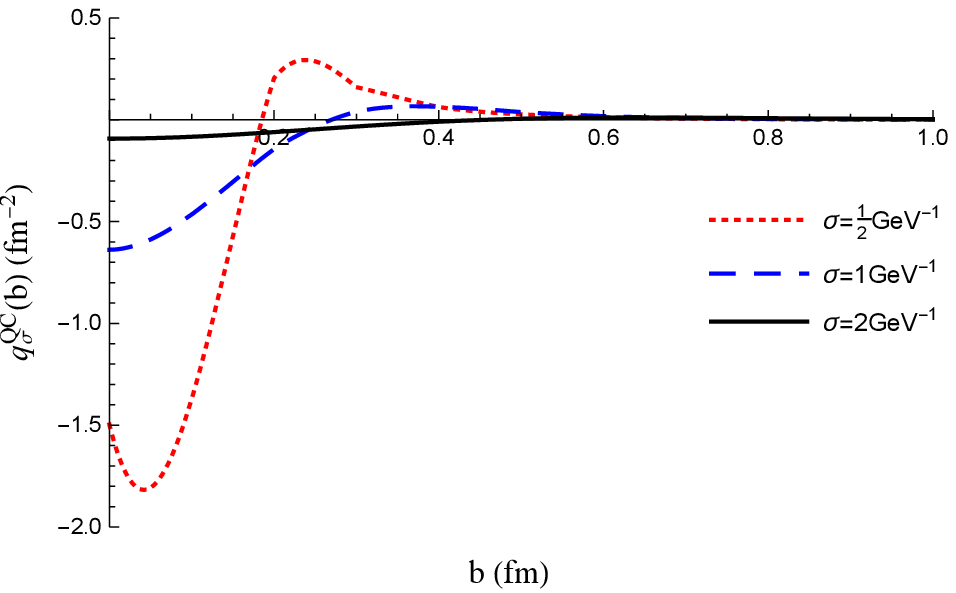}
\end{minipage}
}
\caption{\label{fig:ipPDFCQC}{\small The impact parameter dependent FFs $q^{C}_\sigma(b)$ and $q^{QC}_\sigma(b)$ with $\sigma=1/2~\gev^{-1}$, $1~\gev^{-1}$ and $2~\gev^{-1}$.
}}
\end{figure}

In this work, we simply extend the calculation to the impact parameter dependent PDFs $q(b)$ and impact parameter dependent FFs $q^{C,M,Q}_\sigma(b)$.
Figure~\ref{fig:ipPDF} gives the $q(b)$ with a wave packet, $q_\sigma(b)$, and with a cutoff on the momentum transfer, $q(b,\Delta_0)$, respectively.
The comparison shows that the cutoff ($\Delta_0$) has a similar effect as the wave packet with width $\sigma\sim1/\Delta_0$.
Of course, we expect that the prediction of the constituent quark model is reasonable only in the region of $|t|^{1/2}\leq 2~\gev$ and when the momentum transfer becomes larger the constituent quark model fails.
The width of the wave packet is also constrained by the uncertainty principle: to have a valid probability interpretation of the initial and finial states, the position dispersion ($\sim\sigma$) cannot be smaller than the Compton wavelength. In the later content, our numerical results in Fig.~\ref{fig:ipPDFCQC:C} agree with this point of view. \\

\end{multicols}

\begin{multicols}{2}

\begin{figure}
\centering
{\hskip -1.5cm}
\subfigure[~$q_\sigma(b)$ (fm${}^{-2}$) with packet width $\sigma=1~\gev^{-1}$.]{\label{fig:ipPDFc:s2}
\begin{minipage}[b]{0.4\textwidth}
\includegraphics[width=1\textwidth]{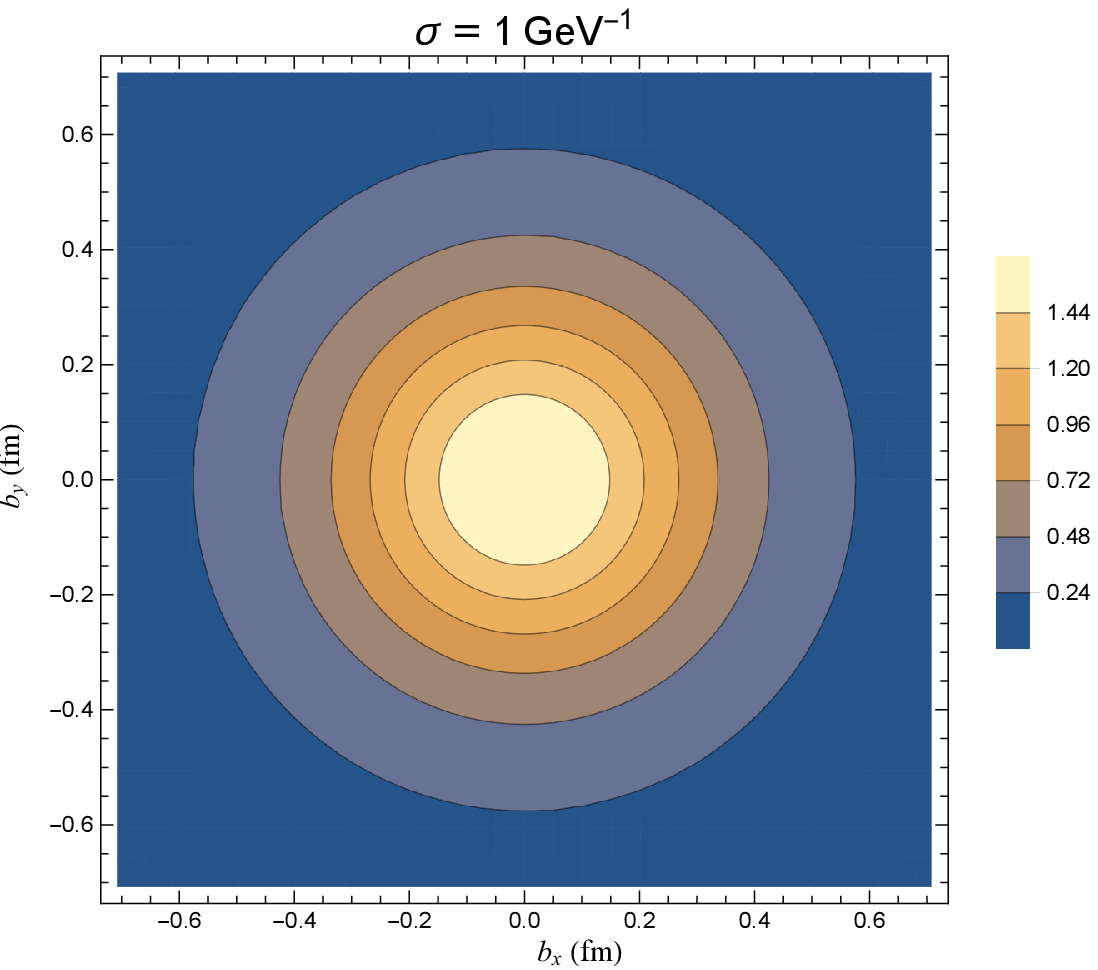}
\end{minipage}
}
{\hskip 1cm}
\subfigure[~$q_\sigma(b)$ (fm${}^{-2}$) with packet width $\sigma=2~\gev^{-1}$.]{\label{fig:ipPDFc:s4}
\begin{minipage}[b]{0.4\textwidth}
\includegraphics[width=1\textwidth]{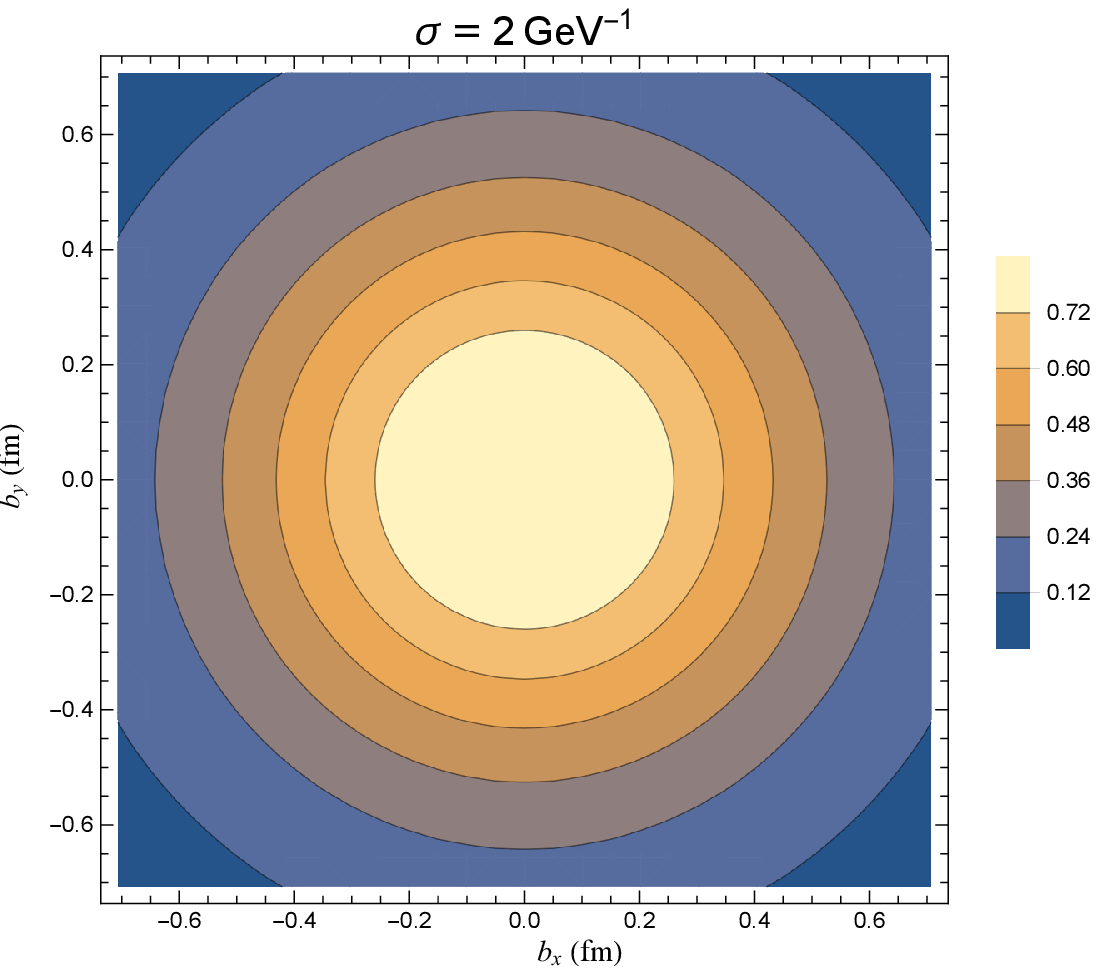}
\end{minipage}
}
\caption{\label{fig:ipPDFc}{\small Contour plots of the impact parameter dependent PDF $q(b)$ with a wave packet.
}}
\end{figure}

Figure~\ref{fig:ipPDFc} gives the contour plots of the impact parameter dependent PDF $q_{\sigma}(b)$  with $\sigma=1~\gev^{-1}$ and $2~\gev^{-1}$.
Since  we choose the polarization in the $z$ direction, the parton distribution is invariant under  rotation around the $z$ direction.
We see that as  $\sigma$ becomes smaller, the wave functions of the initial and final states get  closer to a plane wave, and the parton distribution also becomes more transversely localized in the position space, as shown in Fig.~\ref{fig:ipPDF} and Fig.~\ref{fig:ipPDFc}. \\

\end{multicols}
\begin{multicols}{2}

\begin{figure}
\centering
{\hskip -1.5cm}
\subfigure[~$q_\sigma^M(b)$ with packet width $\sigma=1/2~\gev^{-1}$, $1~\gev^{-1}$, and $2~\gev^{-1}$.]{\label{fig:ipPDFMQ:M}
\begin{minipage}[b]{0.4\textwidth}
\includegraphics[width=1\textwidth]{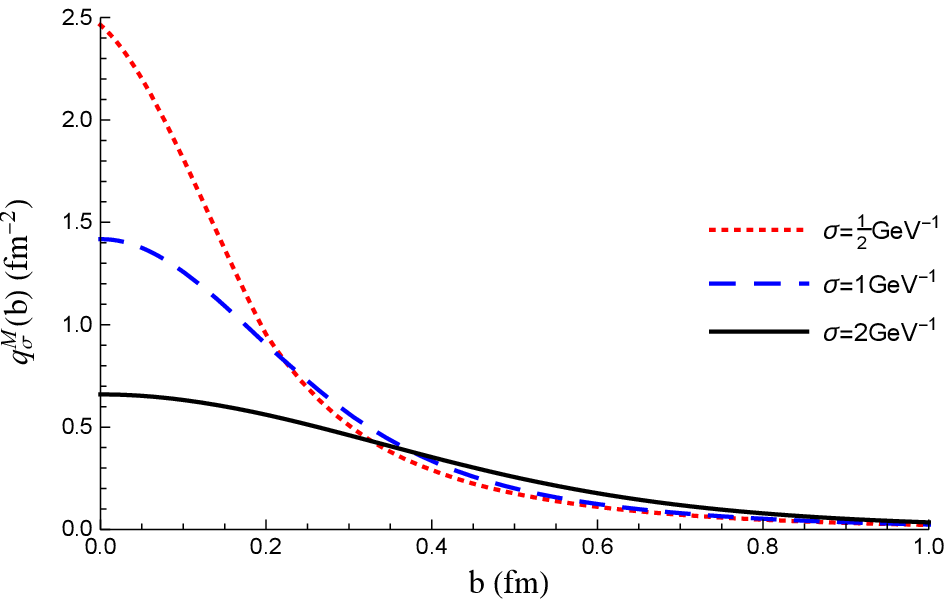}
\end{minipage}
}
{\hskip 1cm}
\subfigure[~$q_\sigma^Q(b)$ with packet width $\sigma=1/2~\gev^{-1}$, $1~\gev^{-1}$, and $2~\gev^{-1}$.]{\label{fig:ipPDFMQ:Q}
\begin{minipage}[b]{0.4\textwidth}
\includegraphics[width=1\textwidth]{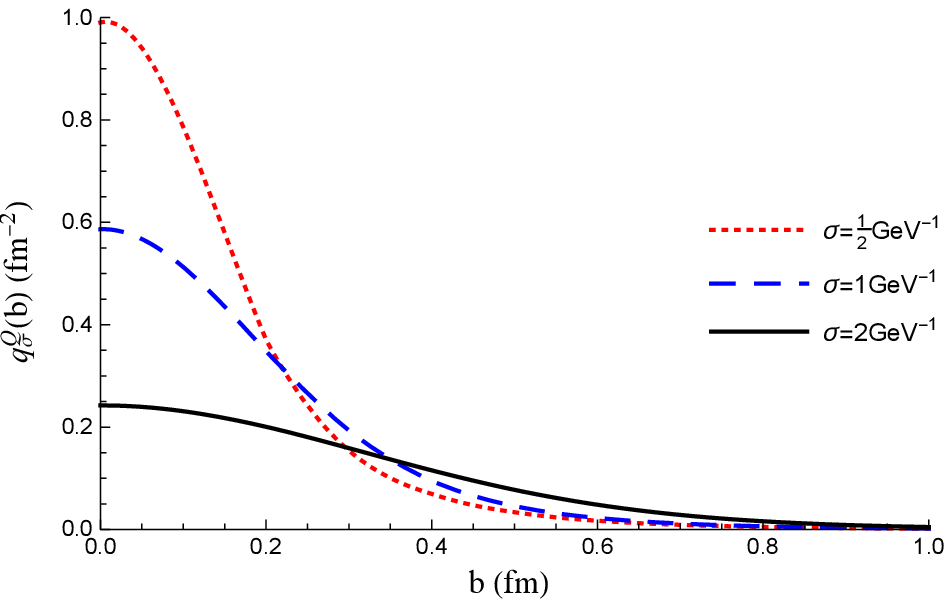}
\end{minipage}
}
\caption{\label{fig:ipPDFMQ}{\small The impact parameter dependent FFs $q^{M,Q}_\sigma(b)$ with $\sigma=1/2~\gev^{-1}$, $1~\gev^{-1}$, and $2~\gev^{-1}$.
}}
\end{figure}

Figures~\ref{fig:ipPDFCQC} and \ref{fig:ipPDFMQ} give the impact parameter dependent FFs $q^{C,M,Q}_\sigma(b)$ and $q^{QC}_\sigma(b)$ with $\sigma=1/2~\gev^{-1},\; 1~\gev^{-1},\; 2~\gev^{-1}$ respectively.
Figure~\ref{fig:ipPDFMQ} shows that, as the wave packet becomes more sharply localized ($\sigma$ decreases), the contributions are concentrated more in the small ${\bf b}_\perp$ region for both the magnetic dipole $\mu_{\rho}$ and quadrupole moment $Q_\rho$.
For the impact parameter charge density, Fig.~\ref{fig:ipPDFCQC:C}, the distributions with $\sigma$ less than about $1~\gev^{-1}$ become obscure due to the oscillation.
As we argued before, the $\rho$ meson is an extended object and its Compton wavelength is $1/m_{\rho}=1.3 \gev^{-1}$.
The position dispersion $\langle\Delta x\rangle=\sigma$ in the case of the Gaussian wave packet. The uncertainty principle ($\langle\Delta x\rangle\langle\Delta p\rangle \ge 1/2$ in natural units) gives the constraint that, to maintain the probability interpretation of the states, the position dispersion $\langle\Delta x\rangle$ should not be smaller than the Compton wavelength.
Otherwise, localizing a wave packet to less than its Compton wavelength in size will in general induce various relativistic corrections~\cite{Burkardt:2000za}.
With the help of Figs.~\ref{fig:ipPDF} and \ref{fig:ipPDFCQC:QC}, and Eq.~\ref{eq:sumCQC}, the oscillation in $q^{C}_\sigma(b)$ can be explained as the behaviour of $q^{QC}_\sigma(b)$ which is related to the quadrupole moment.
From the experimental aspect, since the $\rho$ meson quadrupole moment is small, this phenomenon is hard to determine. \\

\end{multicols}

\begin{multicols}{2}

\begin{figure}
\centering
{\hskip -1.5cm}
\subfigure[~$q_\sigma(x,b)$ with $\sigma=1~\gev^{-1}$ and $x=1/10$, $3/10$, and $1/2$.]{\label{fig:ipPDF1CQCx:1}
\begin{minipage}[b]{0.4\textwidth}
\includegraphics[width=1\textwidth]{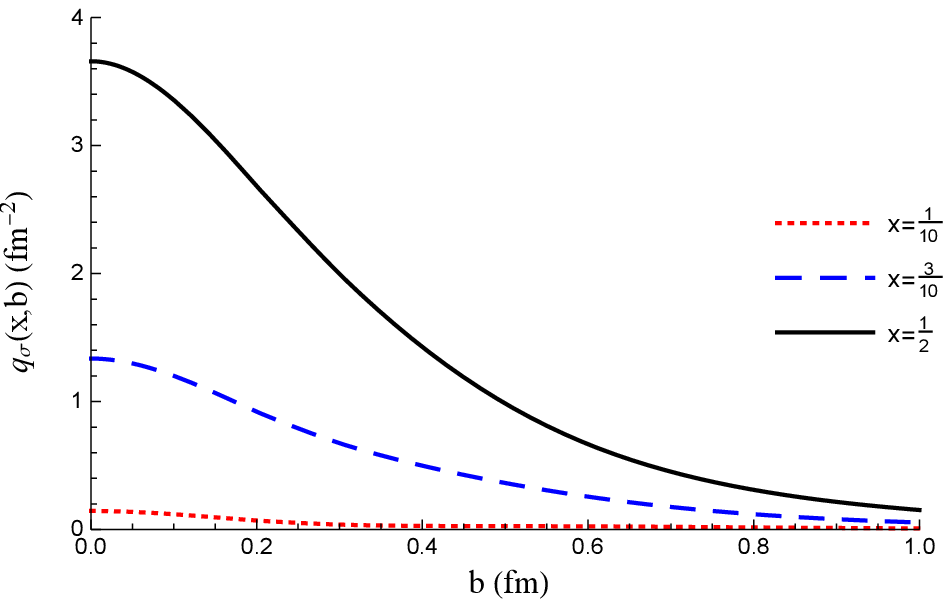}
\end{minipage}
}
{\hskip 1cm}
\subfigure[~$q_\sigma^C(x,b)$ with $\sigma=1~\gev^{-1}$ and $x=1/10$, $3/10$ and $1/2$.]{\label{fig:ipPDF1CQCx:C}
\begin{minipage}[b]{0.4\textwidth}
\includegraphics[width=1\textwidth]{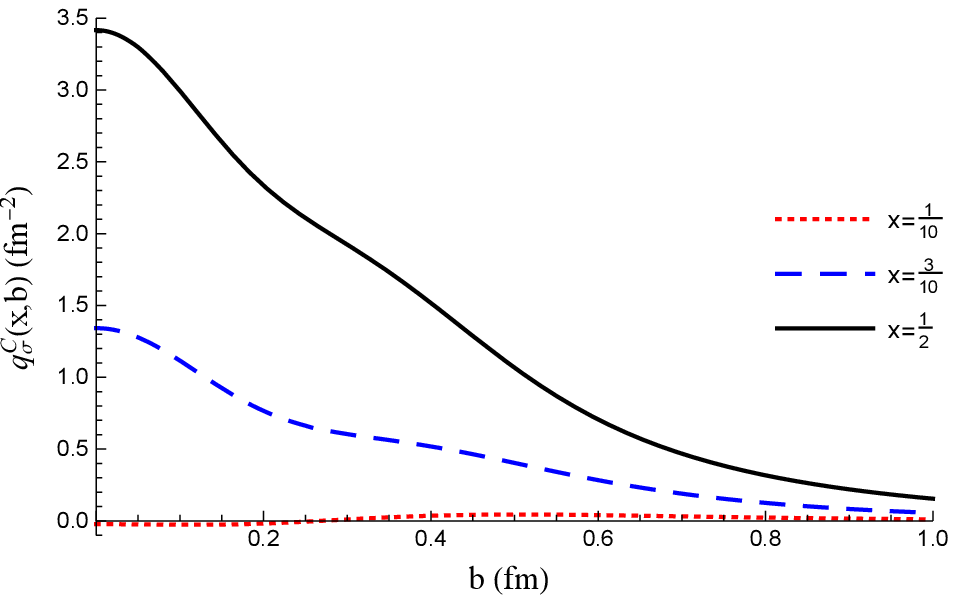}
\end{minipage}
}
\\
{\hskip -1.5cm}
\subfigure[~$q_\sigma^{QC}(x,b)$ with $\sigma=1~\gev^{-1}$ and $x=1/10$, $3/10$ and $1/2$.]{\label{fig:ipPDF1CQCx:QC}
\begin{minipage}[b]{0.4\textwidth}
\includegraphics[width=1\textwidth]{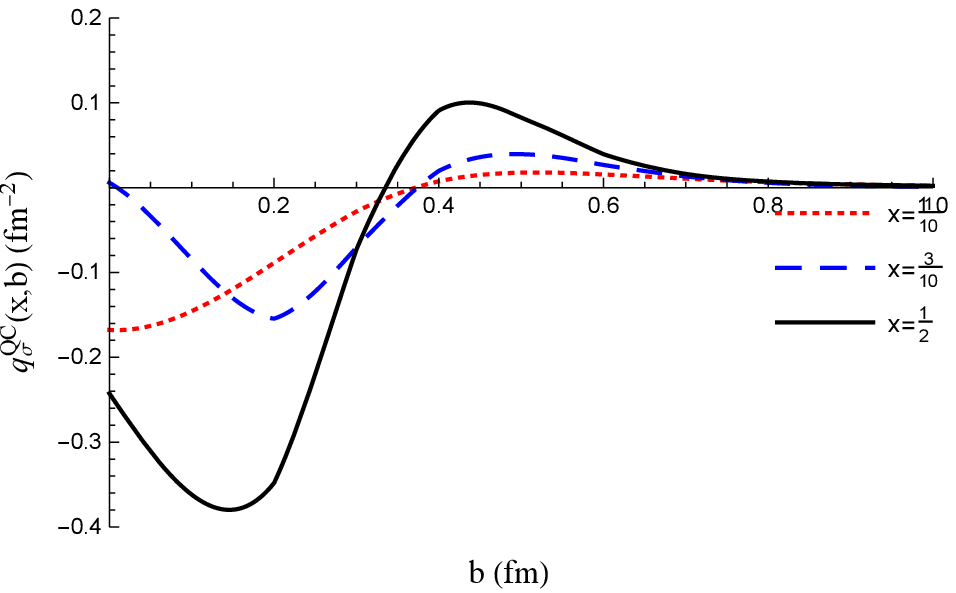}
\end{minipage}
}
\caption{\label{fig:ipPDF1CQCx}{\small The impact parameter dependent PDFs $q_\sigma(x,b)$ and FFs $q^{C,QC}_\sigma(x,b)$ with $\sigma=1~\gev^{-1}$ and $x=1/10$, $3/10$ and $1/2$.
}}
\end{figure}

\begin{figure}
\centering
{\hskip -1.5cm}
\subfigure[~$q_\sigma^M(x,b)$ with $\sigma=1~\gev^{-1}$ and $x=1/10$, $3/10$ and $1/2$.]{\label{fig:ipPDFMQx:M}
\begin{minipage}[b]{0.4\textwidth}
\includegraphics[width=1\textwidth]{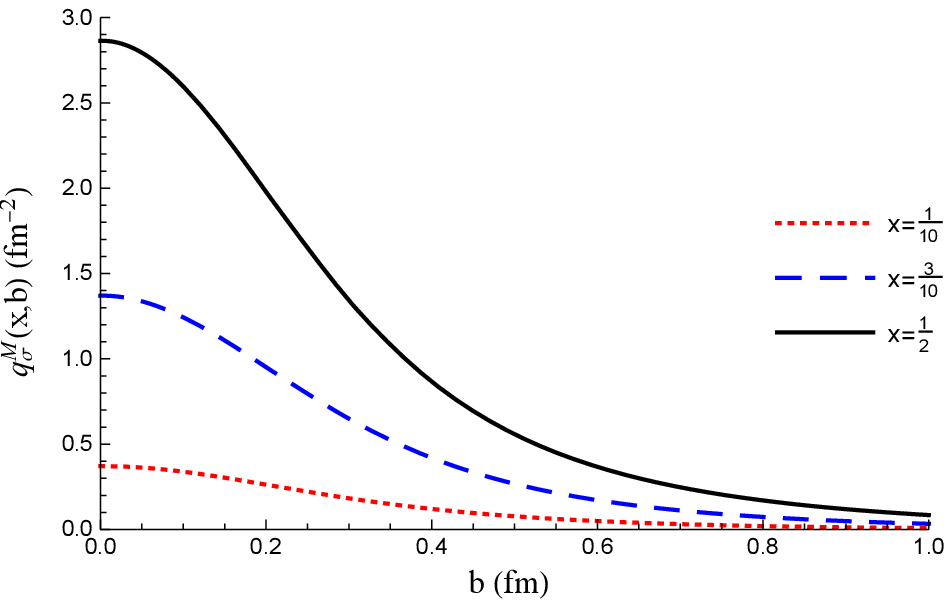}
\end{minipage}
}
{\hskip 1cm}
\subfigure[~$q_\sigma^Q(x,b)$ with $\sigma=1~\gev^{-1}$ and $x=1/10$, $3/10$ and $1/2$.]{\label{fig:ipPDFMQx:Q}
\begin{minipage}[b]{0.4\textwidth}
\includegraphics[width=1\textwidth]{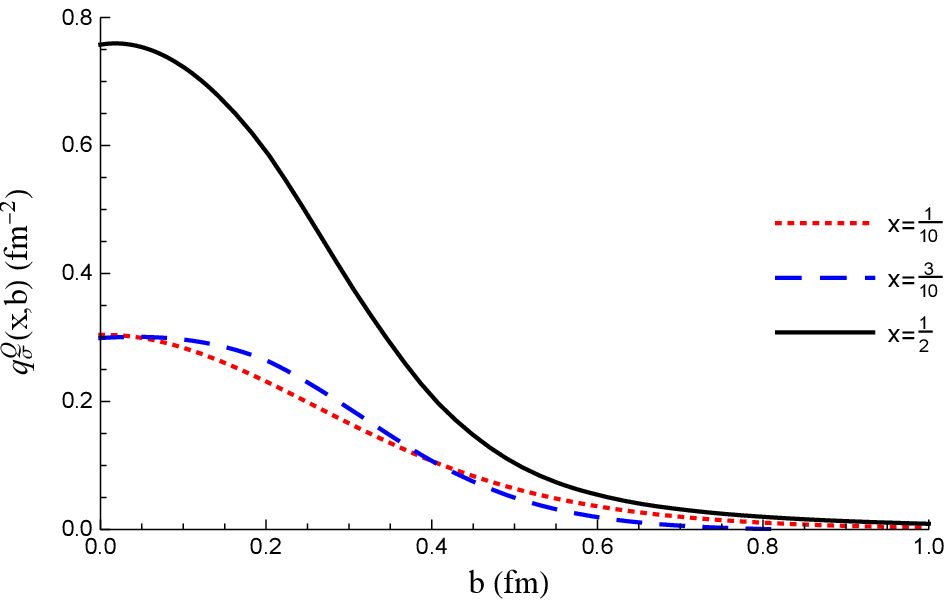}
\end{minipage}
}
\caption{\label{fig:ipPDFMQx}{\small The impact parameter dependent FFs $q^{M,Q,QC}_\sigma(x,b)$ with $\sigma=1~\gev^{-1}$ and $x=1/10$, $3/10$ and $1/2$.
}}
\end{figure}

Figures~\ref{fig:ipPDF1CQCx} and \ref{fig:ipPDFMQx} show the numerical result of $q_\sigma(x,b)$ and $q^{C,M,Q,QC}_\sigma(x,b)$ with $\sigma=1~\gev^{-1}$ and $x=1/10,\;3/10 \; \text{and} \; 1/2$ respectively. When $x~\le~1/10$, $q^{C}_\sigma(x,b)$ has negative values as $b<0.4$~fm (see Fig.~\ref{fig:ipPDF1CQCx:C}), due to the oscillation of $q^{QC}_\sigma(x,b)$ (see Fig.~\ref{fig:ipPDF1CQCx:QC}).
In the small $x$ region (like $x~<~1/10$ in our case), it is believed that the contribution of the gluon GPDs becomes more important, which is beyond the scope of the present model. The symmetry around $x\sim1/2$ of the parton distributions, implied by the isospin symmetry, is not satisfied well due to this reason.
In addition, we found, from  Fig.~\ref{fig:ipPDFMQx}, that the distributions approximately remain the same in $q^Q_\sigma(x,{b})$ when $1/10~\le~x~\le~3/10$.

\end{multicols}

\begin{multicols}{2}

\section{Summary and conclusions}
\label{sec:summary}

In this work, analogous to the definition of the pion and nucleon impact parameter dependent PDFs, we introduce the $\rho$ meson impact parameter dependent PDFs ($q(x,{b})$ and $q({b})$) and impact parameter dependent FFs ($q^{C,M,Q}(x,{b})$ and $q^{C,M,Q}({b})$).
By employing the LCCQ, as we have done previously, we carried out the numerical calculation of those quantities for the first time.
We believe that $q^{C,M,Q}(x, {b})$ may be interpreted as the percentages of the contributions to the charge (normalized to 1), magnetic dipole $\mu_\rho$, and quadrupole moment $Q_\rho$, respectively, from a parton with a longitudinal momentum fraction $x$ and a transverse impact space position ${\bf b}_\perp$.
Considering the facts that the $\rho$ meson is an extended object and there exists a largest measured value of momentum transfer in realistic measurements, a Gaussian form wave packet is employed in our numerical calculation.
Our numerical results show that the wave packet approach plays a similar effect to the cutoff in the integral, which is due to the validity of the constituent quark model.
Our numerical results for impact parameter charge distributions also show that the width of the Gaussian wave packet should be larger than the Compton wavelength.
We expect that this approach is needed in a phenomenological model calculation in order to remove the possible negative values of the impact parameter charge distributions $q^C_{\sigma}(x,{b})$, which cannot be understood by the density interpretation.

\section*{Acknowledgements}
We would like to thank  Stanley J. Brodsky, M. V. Polyakov, and Haiqing Zhou for their encouragement and constructive discussions.
This work is supported by the National Natural Science Foundation of China under Grant No. 11475192, by the fund provided to the Sino-German CRC 110 ``Symmetries and the Emergence of Structure in QCD" project by the NSFC under Grant No.11621131001,  and the Key Research Program of Frontier Sciences,  CAS, Grant No. Y7292610K1.

\end{multicols}
\vspace{2mm}
\centerline{\rule{80mm}{0.1pt}}
\vspace{2mm}
\begin{multicols}{2}

\end{multicols}

\clearpage
\end{CJK*}
\end{document}